\documentclass[aps,prl,twocolumn,groupedaddress,superscriptaddress]{revtex4-1}
\usepackage{amsmath}
\usepackage{amssymb}
\usepackage[hidelinks,colorlinks=true,linkcolor=blue,citecolor=blue,anchorcolor=blue,filecolor=blue,menucolor=blue,runcolor=blue,urlcolor=blue]{hyperref}
\usepackage{bm}
\usepackage{color}
\usepackage{graphicx}
\usepackage{lipsum}

\makeatother

\begin{document}

\title{Floquet topological phase transitions induced by uncorrelated or correlated disorder}
\author{Jun-Hui Zheng}
\email{junhui.zheng@nwu.edu.cn}
\affiliation{Shaanxi Key Laboratory for Theoretical Physics Frontiers, Institute of Modern Physics, Northwest University, Xi'an, 710127, China}
\affiliation{Peng Huanwu Center for Fundamental Theory, Xi'an 710127, China}
\author{Arijit Dutta}
\affiliation{Goethe-Universit{\"a}t, Institut f{\"u}r Theoretische Physik, 60438 Frankfurt am Main, Germany}
\author{Monika Aidelsburger}
\affiliation{Max-Planck-Institut f\"ur Quantenoptik, 85748 Garching, Germany}
\affiliation{Faculty of Physics, Ludwig-Maximilians-Universit\"at M\"unchen, Schellingstra\ss e 4, D-80799 Munich, Germany}
\affiliation{Munich Center for Quantum Science and Technology (MCQST), Schellingstra\ss e 4, D-80799 M\"unchen, Germany}
\author{Walter Hofstetter}
\email{hofstett@physik.uni-frankfurt.de}
\affiliation{Goethe-Universit{\"a}t, Institut f{\"u}r Theoretische Physik, 60438 Frankfurt am Main, Germany}

\begin{abstract}
The impact of weak disorder and its spatial correlation on the topology of a Floquet system is not well understood so far. In this study, we investigate a model closely related to a two-dimensional Floquet system that has been realized in experiments. In the absence of disorder, we determine the phase diagram and identify a new phase characterized by edge states with alternating chirality in adjacent gaps. When weak disorder is introduced, we examine the disorder-averaged Bott index and analyze why the anomalous Floquet topological insulator is favored by both uncorrelated and correlated disorder, with the latter having a stronger effect. For a system with a ring-shaped gap, the Born approximation fails to explain the topological phase transition, unlike for a system with a point-like gap.
\end{abstract}

\maketitle

Topological states are fascinating due to their unique properties and potential applications in spintronics and quantum computation \cite{Hasan2010}. In photonic systems, acoustic systems, and ultracold quantum gases, periodic driving has been extensively employed for engineering topological phases \cite{Weitenberg2021, Jotzu2014, Martin2017, Aidelsburger2015, Rechtsman2013, Eckardt2017, Maczewsky2017, Mukherjee2017, Fleury2016, Peng2016, Wintersperger2020, Rudner2020, Maczewsky2017, Eckardt2015, Qin2018}. In the case of high driving frequency, the evolution of the Floquet system can be effectively described by a stationary Hamiltonian that neglects micromotion on short time scales, providing a mechanism for dynamically realizing topological states \cite{Eckardt2015}. When the driving frequency is comparable to the bandwidth of the system, new features emerge that distinguish the Floquet system from a static one. These features include hybridization effects among different Floquet sectors, resonances during dynamical evolution \cite{Eckardt2017, Qin2018}, and the existence of anomalous Floquet topological insulators (AFTI) with robust edge states but vanishing Chern numbers for all energy bands \cite{Rudner2013, Leykam2016, Maczewsky2017, Mukherjee2017, Fleury2016, Peng2016, Wintersperger2020, Maczewsky2017}.

A driven system in the presence of disorder exhibits even richer behavior, for example, the dynamical many-body localization \cite{Ponte2015, Zhang2016, Khemani2016, Po2016}. Compared to static topological systems where weak disorder induces a phase transition mainly through the sign change of effective masses \cite{Li2009, Groth2009, Zheng2019}, disorder in a Floquet system has multiple effects due to the intrinsic complexity of topological structures. In addition to the greater diversity of disorder-induced band inversion in topological phase transitions \cite{Titum2015, Meier2018, Stutzer2018}, new classes of topological phases are further introduced. For instance, the emergence of a topological Floquet-Anderson insulator where chiral edge modes coexist with a fully localized bulk \cite{Titum2016, Mena2019, Kundu2020, Zheng2022}. Yet,  the mechanism how weak disorder affects the topology of a driven system is still not fully clear, even though strong disorder generally leads to trivial topology. Moreover, in previous studies, disorder potentials on different sites have mostly been assumed to be independent of each other. However, the optical speckle potential in ultracold atom experiments is usually spatially correlated at short distances \cite{Billy2008}. Open questions include which types of topological phases are favored by disorder and what the effect of the spatial correlation of the disorder potential is. In this Letter, we aim to answer these questions.

\begin{figure}
\includegraphics[width=0.99\columnwidth]{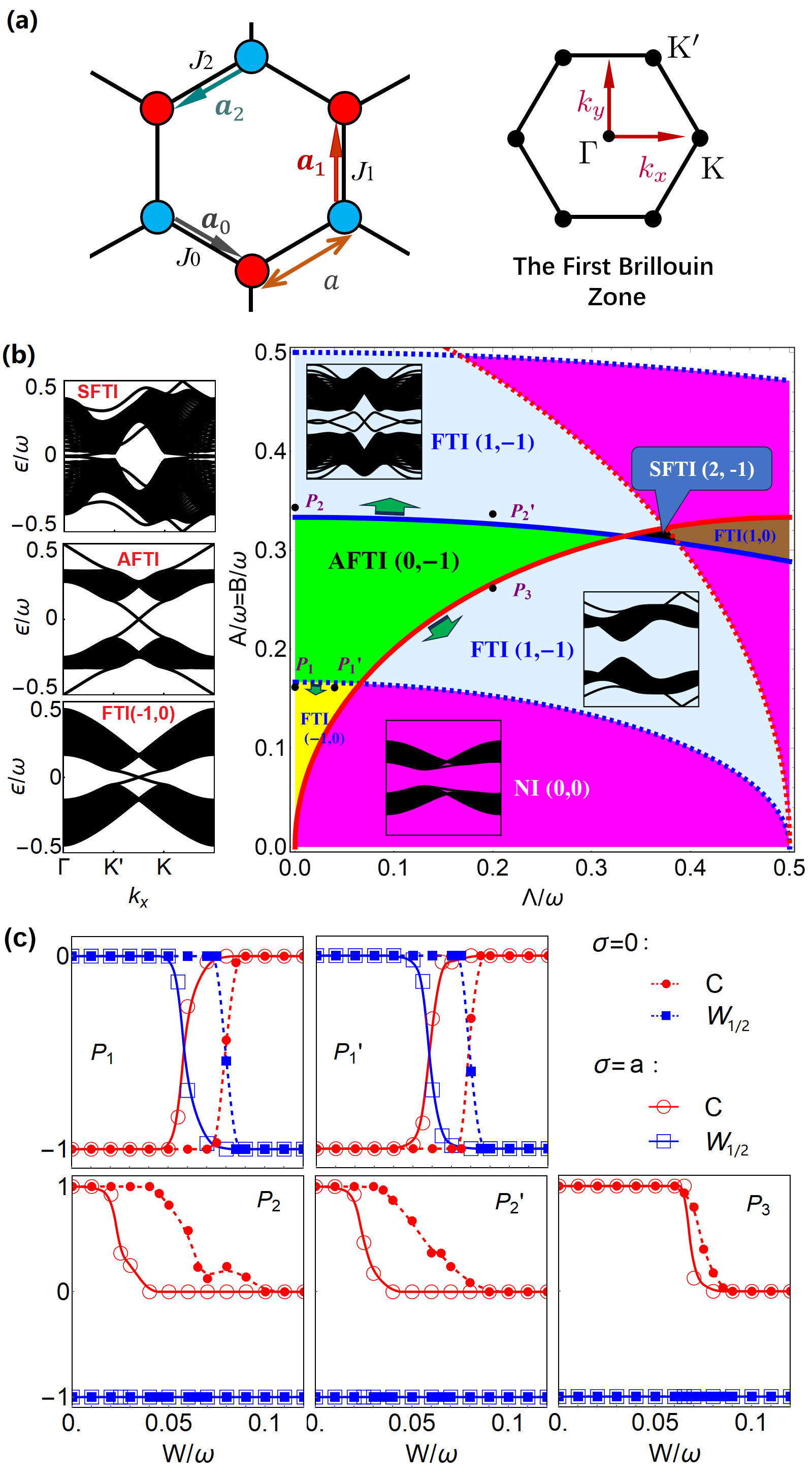}
\caption{(a) Lattice structure in real space and the first Brillouin zone in reciprocal space. (b) The phase diagram for $A=B$ without disorder and the typical spectra. Different topological phases are distinguished by colors. Blue (red) lines correspond to band crossing at the $\Gamma$ point (the Dirac points).  Solid (dashed) lines refer to band crossing at $\epsilon =0~(\omega/2)$. The phase boundary of the AFTI phase is shifted along the directions indicated by the green vectors when disorder is present. (c) Topological phase transition from FTI to AFTI induced by uncorrelated disorder ($\sigma=0$) and correlated disorder ($\sigma=a$) for different points in the FTI phases. All these points become AFTI when the disorder strength is increased. Here, the points  $P_1$, $P'_1$ and $P_3$ correspond to $\Lambda=0$, $\Lambda/\omega=0.04$, and $\Lambda/\omega=0.2$, respectively, and a value of $A/\omega=B/\omega$ that is $0.005$ below the respective phase transition point. For $P_2$ and $P'_2$ parameters are $\Lambda=0$ and $\Lambda/\omega=0.2$, respectively, with a value of $A/\omega=B/\omega$ that is $0.01$ above the respective phase transition point. Numerical calculations have been performed for $N_x =29$ and $N_y =34$ with 30 samples of disorder, where $N_x$ and $N_y$ refer to the numbers of unit cells along $x$ and $y$ direction, respectively.}
\label{figs1}
\end{figure}

We consider a model closely related to the one which has been experimentally realized in a two-dimensional bosonic ultracold atom system with a honeycomb lattice (see Fig.\,\ref{figs1}(a)) \cite{Wintersperger2020}. The coefficients of hopping between neighboring sites in different directions (i.e., ${\bm a}_\lambda/ a$ for $\lambda =0,1,2$, where the ${\bm a}_\lambda$ are vectors between pairs of neighboring sites and $a =|\bm a_\lambda|$), vary periodically with time and reach the maximum value in turn,
\begin{equation}
J_\lambda (t) = A+ {B} \cos(\omega t - \phi_\lambda),
\end{equation}
where the phase $\phi_\lambda = {2\pi} \lambda/3$ modulates the hopping strength. A staggered potential of strength $\Lambda$ is further introduced, with opposite values on the blue and red  sublattices. The dynamics of the driven system can be described by a time-independent Hamiltonian $\mathcal{H}$ in the extended Floquet Hilbert space\,\cite{Eckardt2015,Titum2015}, which consists of block matrices $ \mathcal{H}_{mn} = m \omega \delta_{mn} + \int_0^{2\pi/\omega} dt H(t) e^{i\omega(m-n)t}$. In the absence of disorder, all nonvanishing block matrices in momentum space are
\begin{eqnarray}
&&\mathcal{H}_{mm}({\bf k}) = m\omega {\bf 1}_{2\times 2}  -  A \sum_{\lambda=0}^2 f_\lambda({\bf k}) + \Lambda \sigma_z,  \label{floquetH}\\
&&\mathcal{H}_{m,m\mp 1}({\bf k}) = - \frac{ B}{2} \sum_{\lambda=0}^2 \exp\left( \pm  i \phi_\lambda \right) f_\lambda({\bf k}),
\end{eqnarray}
where $f_\lambda({\bf k}) = \sigma_x  \cos({\bf k}\cdot {\bm a}_\lambda)  +  \sigma_y \sin({\bf k}\cdot {\bm a}_\lambda) $. The above identity matrix $ {\bf 1}_{2\times2}$ and Pauli matrices $\sigma_{x(y,z)}$ act on the sublattice space.

{\it Phase diagram} --- Using a truncated extended Floquet Hilbert space (we choose $|m|,|n| \leq 4$ in our calculation) \cite{Rudner2013}, we calculate the Chern number $\mathcal{C}$ ($- \mathcal{C}$) for the band just below (above) zero quasi-energy and the winding numbers $\mathcal{W}_0, \mathcal{W}_{1/2}$ at different quasi-energies $\epsilon = 0, \omega/2$. The winding number, which equals the total Chern number of bands below the given energy in the truncated Floquet Hilbert space, indicates the number of robust edge states in the gap, and the sign of this number determines their chirality. The two winding numbers satisfy $\mathcal{W}_0 = \mathcal{W}_{1/2} + \mathcal{C}$. Different topological phases are classified by indices $(\mathcal{C}, \mathcal{W}_{1/2})$.

\noindent

In Fig.\,\ref{figs1}(b), we plot the phase diagram in the $A$-$\Lambda$ plane for $A=B$ and the typical spectra in different phases: normal insulator (NI) with indices $(\mathcal{C}, \mathcal{W}_{1/2}) = (0, 0)$, Floquet topological insulator (FTI) with $(\pm 1, 0)$ or $(1, -1)$, AFTI with $(0, -1)$, and staggered Floquet topological insulator (SFTI) with $(2, -1)$. In FTI, only one winding number is nonzero, and edge states exist in the corresponding gap.  In AFTI, both winding numbers are the same and nonzero.  Compared to AFTI, the phase SFTI $(2,-1)$ has $\mathcal{W}_{0,1/2}= \pm 1$ and thus edge states in neighboring gaps have opposite chirality. This newly found phase has different transport properties and no clear chirality can be observed from the time evolution of wave-packets at the edge \cite{sm,Martinez2023}. Note that in this model, topological phase transitions occur when the energy gap at the $\Gamma$ point or the Dirac points K or K$'$ closes. In the phase diagram, blue and red lines denote the phase boundaries due to band crossing at the $\Gamma$ point and the Dirac points, respectively; solid (dashed) lines indicate that the band crossing occurs at $\epsilon=0$ ($\omega/2$).

Starting from small $A, B$ (i.e., the high-frequency case) and $\Lambda=0$, the effective stationary Floquet Hamiltonian \cite{Eckardt2015} is given by $H_{F} = \mathcal{ H}_{00}({\bf k})$  $-({\sqrt{3} {B}^2}/{2\omega}) \sum_{\lambda=1}^3\sin({\bf k}\cdot {\bf b}_\lambda) \sigma_z $, where ${\bf b}_\lambda = {\bm a}_\lambda - {\bm a}_{\lambda+1}$ with ${\bm a}_4 \equiv {\bm a}_1$. The second term arises from the first-order correction contributed by couplings between the Floquet sectors $m=0$ and $\pm 1$.  $H_F$ is exactly the Haldane model and the correction term represents an effective path-dependent hopping $\pm i{\sqrt{3} B^2}/{4\omega}$ between next nearest neighbors  \cite{Haldane1988}. When increasing the staggered potential, the system undergoes a phase transition from FTI $(-1,0)$ to NI $(0,0)$ at $(B/\omega)^2= 4 \Lambda / 9\omega$ (band inversion occurs at the Dirac points at $\epsilon =0$). On the other hand, the AFTI emerges when $A = B > \omega/6$, after the band inversion at the $\Gamma$ point at $\epsilon =\omega/2$. Further increasing $A$ and $B$ to $\omega/3$, band inversion occurs at the $\Gamma$ point at $\epsilon =0$. The system then goes into the FTI $(1,-1)$ phase, where only the edge state in the gap at $\epsilon =\omega/2$ remains stable.

Note that at the $\Gamma$ point, $\mathcal{H}_{mm} = m \omega \sigma_0 -3 A \sigma_x + \Lambda \sigma_z$ and
$\mathcal{H}_{m,m\pm1} = 0$. The band touching occurs when the spectra coincide with $0$ or $\omega/2$,
\begin{equation}\label{aboundary}
m\omega \pm \sqrt{\Lambda^2 +9 A^2} = 0 \ \text{ or }\ \frac{\omega}{2}.
\end{equation}
At the Dirac points, $\mathcal{H}_{mm} = m \omega \sigma_0 + \Lambda \sigma_z $ and $\mathcal{H}_{m,m-1} =-({3B}/{2}) \sigma_{+} $ and $-({3B}/{2}) \exp({i2\pi/3}) \sigma_{-}$,
where $\sigma_{\pm} = (\sigma_x \pm i \sigma_y)/2$. The band gap closes when
\begin{equation} \label{bboundary}
\left(m-\frac{1}{2}\right)\omega \pm \sqrt{\left(\frac{\omega}{2}\pm\Lambda\right)^2 + \frac{9 B^2}{4}} = 0 \text{ or } \frac{\omega}{2}.
\end{equation}
Equations\,\eqref{aboundary} and \eqref{bboundary} determine the phase boundaries shown in the phase diagram. The parameters $A$ and $B$ can be respectively used to control the topological phase transition through gap closings at the $\Gamma$ point and the Dirac points for a given value of $\Lambda$.

{\it Phase transitions induced by disorder} --- When disorder is present, the block matrices in real space, $\mathcal{H}_{mn}$, obtain an additional contribution $\delta \mathcal{H}_{mn}^{ij}= \delta_{mn}\delta_{ij}V_{i}$ from the on-site disorder, where $i$ and $j$ denote lattice sites. The optical speckle disorder \cite{Kondov2011,Jendrzejewski2012} is correlated in real continuous space,
\begin{equation}\label{discorr}
    \langle V({\bf r}) V ({\bf r'}) \rangle  = W^2 \exp\left[- \frac{|{\bf r}- {\bf r}'|^2}{2 \sigma^2}\right],
\end{equation}
where $ 2\sigma$ is the correlation length of the disorder. The disorder at different sites becomes uncorrelated when $\sigma \ll a$. To numerically generate samples of correlated disorder, we employ the Fourier transformation, $ V_{\bf k}= \frac{1}{2\pi}\int d{\bf r} V({\bf r}) \exp[{i {\bf k} \cdot {\bf r}}]$. Since $V({\bf r})$ is real, we have $V_{\bf k} =V_{-\bf k}^*$. Denoting $ V_{\bf k} \equiv u_{\bf k} + i v_{\bf k}$ with real variables, eq.\,\eqref{discorr} gives $\langle u_{\bf k} u_{\bf k'}\rangle = \langle v_{\bf k} v_{\bf k'}\rangle = D_{k}^2  \delta_{{\bf k}{\bf k}'} $ for ${\bf k}$ {or} ${\bf k'}\neq 0$, $\langle u_{0} u_{0}\rangle = 2D_{0}^2$,  $\langle v_{0} v_{0}\rangle =0$, and $\langle u_{\bf k} v_{\bf k'}\rangle = 0$, where $D_k = \frac{\sqrt{\pi}W\sigma}{\sqrt{\Delta k_x \Delta k_y}} \exp(-{\sigma^2 {\bf k}^2}/{4})$, and $\Delta k_\alpha$ is the discrete spacing in momentum space for $\alpha=x,y$. Thus, $u_{\bf k}$ and $v_{\bf k}$ are independent random variables. For periodic boundary conditions, we further require $\Delta k_\alpha = 2\pi/L_\alpha$ and $k_\alpha$ being a multiple of $\Delta k_\alpha$ to make the disorder correlated in a torus geometry, where $L_x$ and $ L_y$ are the length and width of the system, respectively.  In our numerical simulation, we choose a uniform distribution $u_{\bf k}, v_{\bf k} \in [-\sqrt{3}D_k , \sqrt{3}D_k]$ for ${\bf k}\neq 0$,  and $u_{0}\in[-\sqrt{6}D_0, \sqrt{6}D_0]$ to achieve the corresponding variances $D_k^2$ and $2D_{0}^2$. Given a random sample of $u_{\bf k}$ and $v_{\bf k}$, the disorder $V({\bf r})$ is obtained by the inverse Fourier transformation.

In disordered systems, the relevant  topological index is the Bott index \cite{Loring2011}, $\mathcal{B} = (1/2\pi) \text{Im[Tr[log}[ \hat U_x \hat U_y \hat U^\dagger_x \hat U^\dagger_y]]]$, where $U_\alpha = \hat P \exp(-2i\pi \hat{\alpha}/L_\alpha) \hat P$ for $\alpha =x,y$ with $P$ being the projection operator to the state subspace of a given band (this corresponds to the Chern number) or of the bands below some energy (this corresponds to the winding number). Since $\hat x$ and $\hat y$ are the generators of the translation operators in momentum space, the Bott index gives the `magnetic' flux in momentum space, which recovers the Chern number or the winding number when $W \rightarrow 0$. The disorder-averaged Bott indices show that the states in regions near the AFTI phase will transit to AFTI when increasing the disorder strength, thus broadening the parameter region of the AFTI phase, as is sketched in Fig.\,\ref{figs1}(b). How topological indices $ (\mathcal{C}, \mathcal{W}_{1/2})$ change with the disorder strength is shown in Fig.\,\ref{figs1}(c) for different points outside of the AFTI phase. With increasing the correlation length $\sigma$ from $0$ to $a$, the phase transitions occur at smaller disorder strength. These results indicate that AFTI is favored by uncorrelated disorder and even more by correlated disorder.

{\it  Analysis and discussion} --- Usually, the effect of weak disorder is interpreted via the effective self-energy using the self-consistent Born approximation \cite{Groth2009,Zheng2019,Titum2015}. For the time-independent Hamiltonian $\mathcal{H}$ in the extended Floquet Hilbert space this self-energy takes the following form, written in real space at the quasienergy $\epsilon$:
\begin{equation}\label{self}
    \Sigma_{mn}^{ij}(\epsilon) = G_{mn}^{ij}(\epsilon) \langle V_i V_j  \rangle,
\end{equation}
where $G(\epsilon)= 1/(\epsilon-\mathcal{H}-\Sigma)$ is the Green's function and the indices $i$, $j$ denote lattice sites \cite{Zheng2019}. This self-energy correction is an approximation which usually is valid for extended (delocalized) states at weak disorder. The topological phase transition occurs when the mobility gap of the system (which can be obtained from the Green's function) closes. When we focus on the winding number $\mathcal{W}_0$, the corresponding reference energy is $\epsilon =0$; while for $\mathcal{W}_{1/2}$ it is $\epsilon =\omega/2$. From Eq.\eqref{self}, we see that for uncorrelated disorder, $\Sigma_{mn}^{ij}(\epsilon)$ vanishes for $i\neq j$, while for correlated disorder, it can be nonzero, representing corrections to hopping coefficients.

We first consider the disorder-induced phase transition in the region where $P_1$ is located [see Fig.\,\ref{figs1}(b)]. The self-energy restores the translational symmetry in the Born approximation. Since the phase transition occurs at the $\Gamma$ point during band inversion at $\epsilon = {\omega}/2$, we focus on the self-energy at this point, which is given by
\begin{equation}
\Sigma_{mn}^{ss'}\left(\frac{\omega}{2}\right)
     \simeq  \frac{3\sqrt{3}a^2 W^2}{8 \pi^2}  \int_{\text{BZ}} d{\bf k} \left[\frac{1}{{\omega}/{2}-\mathcal{H}({\bf k})}\right]_{mn}^{ss'} \label{self3}
\end{equation}
for the uncorrelated disorder case, and
\begin{equation}
\Sigma_{mn}^{ss'}\left(\frac{\omega}{2}\right)
     \simeq  \frac{W^2 \sigma^2}{2\pi}  \int d{\bf k} e^{- \frac{1}{2}\sigma^2|{\bf k}|^2} \left[\frac{1}{{\omega}/{2}-\mathcal{H}({\bf k})}\right]_{mn}^{ss'} \label{self2}
\end{equation}
for $\sigma\gtrsim a$, where $s, s'=0,1$ are sublattice indices. The self-energy on the right side (the Green's function) of Eq.\eqref{self} has been neglected \cite{Groth2009}.
To get insight into the mechanism of disorder-induced phase transition, we develop a low-energy effective theory to simplify $\mathcal{H}({\bf k})$ near the $\Gamma$ point, for the case $\Lambda =0$. By employing the unitary transformation $U= \otimes_{m} \exp[-i{\pi}\sigma_{y}/4]$ to rotate both $\mathcal{H}$ and $\Sigma$ in the sublattice space for Eq.\,\eqref{self3} or \eqref{self2}, we finally obtain a low-energy effective $2\times 2$ Hamiltonian \cite{sm}, $\mathcal{H}({\bf k})-{\omega}/{2} \simeq \begin{pmatrix}
M_0& M_1 \\
M_1^* & -M_0
\end{pmatrix}$, where $M_0({\bf k}) = {\omega}/{2} -3A + {3}A a^2 {\bf k}^2/{4}$,  and $M_1({\bf k}) = {3 B a} (1 -\sqrt{3}i) (k_x+ik_y)/{8}$.

\begin{figure}
  \centering
  \includegraphics[width=\columnwidth]{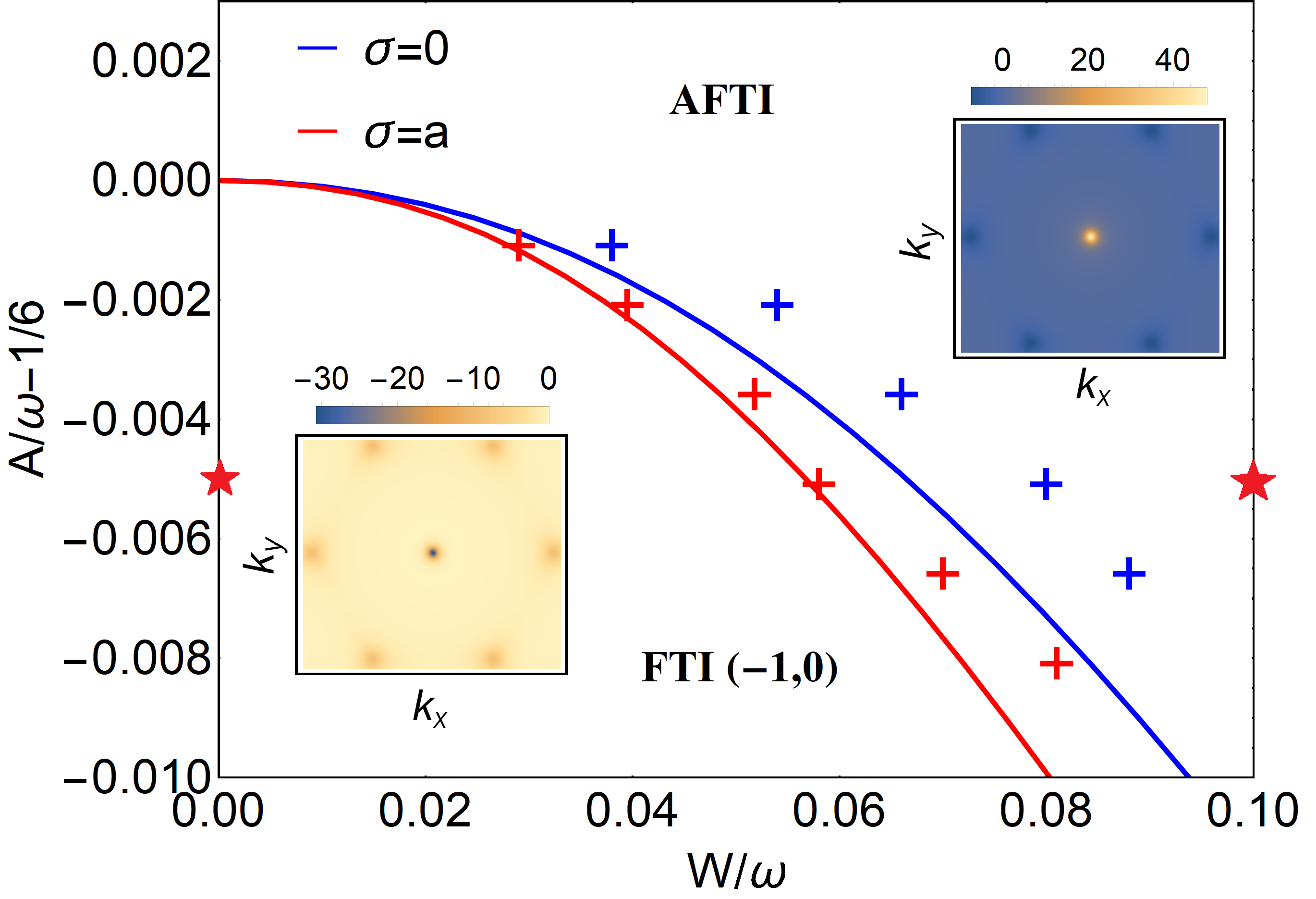}\\
  \caption{The phase transition boundary (solid lines) within the Born approximation for $A=B$ and $\Lambda =0$ in the framework of the low-energy effective model. The points marked with $+$ are determined by the disorder-averaged Bott indices. Inset are the Berry curvatures in momentum space for the two points marked with stars. For $W/\omega=0.1$, it is evaluated in the Born approximation for $\sigma =0$.
  }\label{figs2}
\end{figure}

The obtained effective model is quite similar to the low-energy description of a HgTe quantum well \cite{Groth2009,Bernevig2006}. An effective staggered potential $3\omega(1/6 - A/\omega)>0$ appears in the rotated sublattice space (see $M_0$). This staggered potential is suppressed by the self-energy contributed by disorder \cite{Groth2009}, inducing a topological phase transition by changing its sign (i.e., increasing $A/\omega$). The corresponding phase boundary in the $A$-$W$ plane is shown in Fig.\,\ref{figs2}. Results from the Born approximation in the low energy effective theory quantitatively agree with numerical results obtained via the disorder-averaged Bott index. The critical strength of disorder for the transition is smaller for a larger correlation length. We can also interpret these results from Eq.\,\eqref{self} directly. $\omega$ is an energy bias between neighboring Floquet sectors. This bias gets suppressed by disorder, i.e., the sign of $\sum_{i \in \text{all~sites}}(\Sigma_{11}^{ii}-\Sigma_{00}^{ii})$ is negative \cite{sm}, and thus $A/\omega$ is effectively increased. This explains why the point $P_1$ transits from FTI to AFTI in the presence of uncorrelated disorder. When the correlation length of the disorder is finite, the resulting finite self-energy $\Sigma_{00}^{ij}$ for neighboring $i$ and $j$ contributes an additional correction to the hopping strength $A$, which further effectively increases $A/\omega$ and leads to the phase transition \cite{sm}. In Fig.\,\ref{figs2}, the Berry curvatures for the clean and disordered systems ($\mathcal{H}_{\text{eff}} = \mathcal{H}+\Sigma$ in the Born approximation) are also present. The value at the $\Gamma$ point is significantly changed by disorder.

For point $P'_1$, the staggered potential $\Lambda$ will also be renormalized by disorder \cite{Titum2015}. However, since the phase transition from FTI $(-1,0)$ to AFTI only weakly depends on the change of $\Lambda$ compared to that of $A$ [see Eq.\eqref{aboundary} and Fig.\,\ref{figs1}(b)], the disorder-induced phase transitions at points $P_1$ and $P'_1$ are quite similar [see Fig.\,\ref{figs1}(c)] \cite{sm}.

For point $P_3$, the phase transition occurs when the band gap at the Dirac point closes. The transition is mainly driven by the suppression of  $\Lambda$ and $\omega$ due to the self-energy induced by the disorder \cite{sm}. Note that the correction of the hopping strength $A$ will not change this phase boundary (see Eq.\eqref{bboundary}). Therefore, the effect of disorder correlation is primarily a renormalization of the parameter $B$ contributed by $\Sigma_{m,m\pm 1}$ \cite{sm}.

\begin{figure}[tpb]
\centering
\includegraphics[width=\columnwidth]{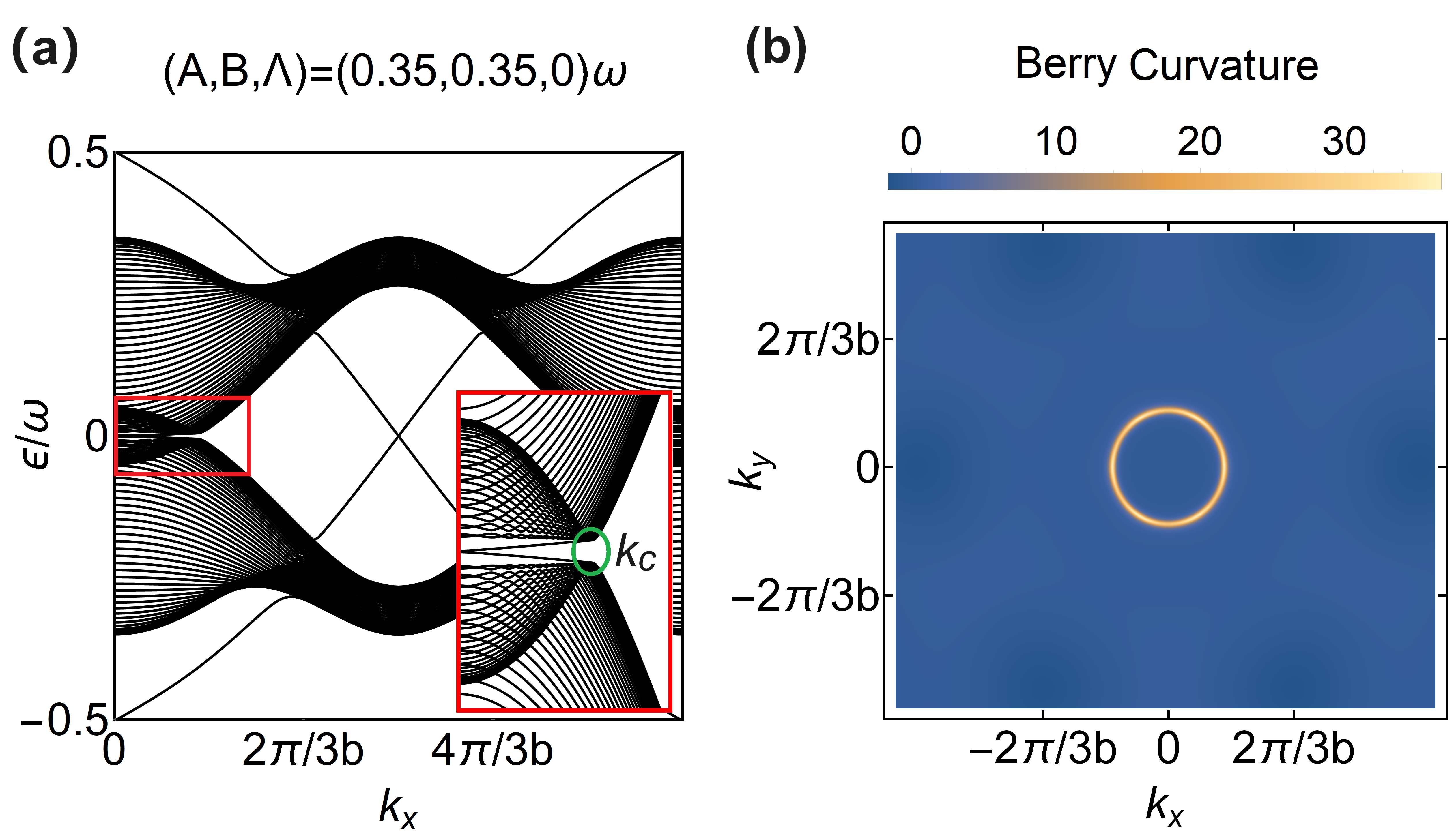}
\caption{(a) The quasi-energy spectrum of the clean system for $A=B = 0.35\omega$ and $\Lambda=0$ in the FTI $(1,-1)$ phase and (b) the corresponding Berry curvature for the lower band. Here, $b= \sqrt{3}a$. The inset figure in (a) shows the enlargement of the red box part, which is the typical spectrum structure with a ring gap at $\epsilon = 0$ for the parameter region just above the upper phase boundary (the blue solid line) of AFTI shown in Fig.\,\ref{figs1}(a). The band gap near $\epsilon =0$ arises from the coupling between different Floquet sectors (see the green ellipse).}
\label{figs3}
\end{figure}

The disorder-induced phase transition for the region just above the upper boundary of the AFTI is unconventional. The typical spectrum for a clean system in this FTI $(1,-1)$ region and the corresponding Berry curvature are shown in Fig.\,\ref{figs3}. The hybridization between different Floquet sectors opens a ring gap at $\epsilon =0$ in the Brillouin zone for $|{\bf k}| = k_c$. The value $k_c$ depends on the magnitudes of $A$, $B$, and $\Lambda$. The edge states in this gap are not stable ($\mathcal{W}_0 = 0$). The weak on-site disorder couples the degenerate and near-degenerate states near the ring and induces the recombination of these states. Therefore, increasing the disorder strength violates the Berry curvature structure near the ring and erases the nonzero Chern number. As a result, the winding number in the gap  at $\epsilon =0$ becomes the same as the one at the neighbouring gap with $\mathcal{W}_{1/2} =-1$,  leading to the phase transition from FTI to AFTI. In this case, the mobility gap closes at the ring, which cannot be described by an effective Hamiltonian with self-energy corrections from the Born approximation \cite{sm,Titum2015} (note that by changing the system parameters $A$, $\omega$, and $\Lambda$, the topological phase transition occurs through closing the gap at the $\Gamma$ point).  Moreover, the spatial correlation of disorder enhances the hybridization among the degenerate states at the ring, resulting in a shift of the phase transition point.

{\it Conclusion} --- We have investigated the phase diagram of an experimentally relevant two-dimensional Floquet system and analyzed how weak disorder and its spatial correlation impact the phase boundaries between different topological phases. A novel phase with edge states of alternating chirality in neighboring gaps has been found. In addition, the AFTI phase is surrounded by different FTI phases. In the parameter region near AFTI, the FTI phases generally go into AFTI with increasing disorder strength. For a point-like gap, the phase transition can be interpreted by renormalizing the system's parameters using the Born approximation, where the Berry curvature at the $\Gamma$ point or Dirac points is significantly changed. The correlation of disorder further enhances this effect through the correction of hopping coefficients. For a ring-shaped gap \cite{Salerno2020}, the Born approximation does not work. The Berry curvature structure near the ring is destroyed by disorder, leading to the transition from FTI to AFTI.

\begin{acknowledgments}
This work was supported by  the NSFC under Grant No.12247103, Shaanxi Fundamental Science Research Project for Mathematics and Physics under Grant No.\,22JSQ041, the Deutsche Forschungsgemeinschaft (DFG, German Research Foundation) under Project No. 277974659 via Research Unit FOR 2414, and by the DFG under Germany's Excellence Strategy - EXC - 2111 - 3908148. This work was also supported by the DFG via the high performance computing center Center for Scientific Computing (CSC).
\end{acknowledgments}

~\\
\newpage

~\\
\newpage

{\bf \Large Supplementary materials}\\

This supplement contains three sections. In the first section, the dynamic evolutions of a wave packet localized at the zigzag edge are present for AFTI and SFTI, respectively. In the second section, the low-energy two-level model is developed, and the disorder-induced topological phase transition is discussed. In the last section, the effective parameters in the Born approximation are summarized. We show that the Born approximation gives a different result from the disorder-averaged Bott index for the case with a ring-shaped gap.

\section{I. Difference between SFTI and AFTI}
\begin{figure}[t]
\includegraphics[width=0.96\columnwidth]{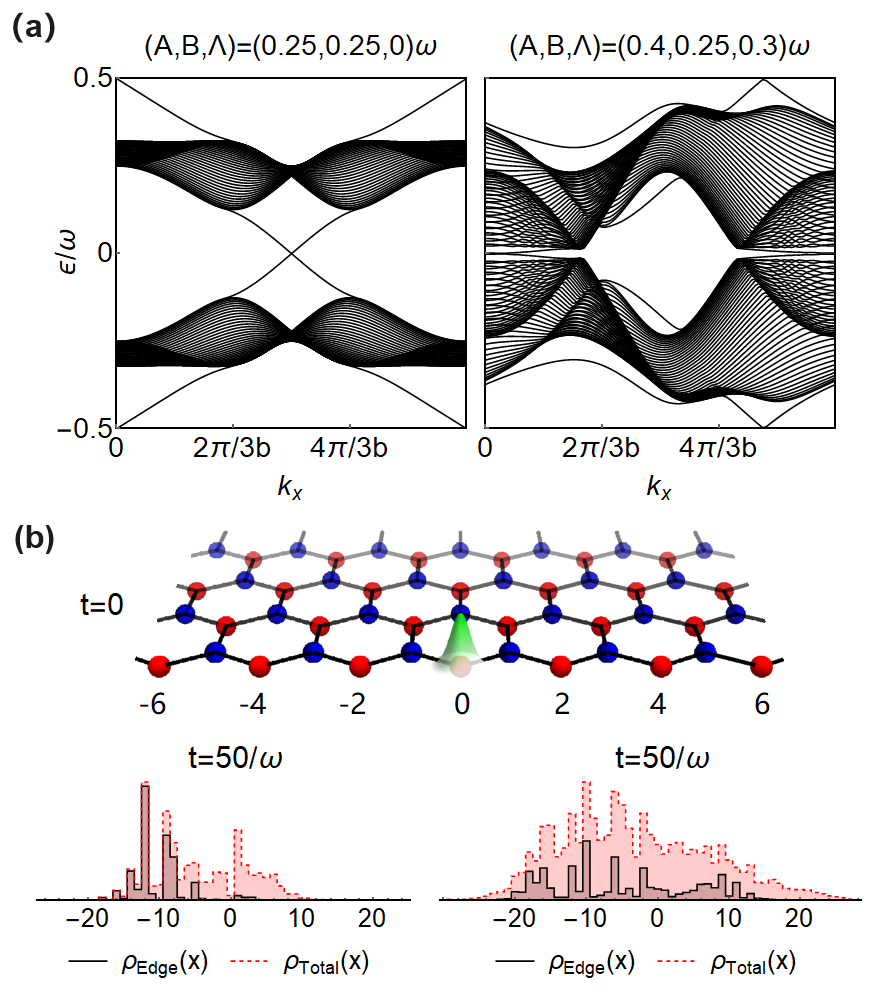}
\caption{(a) The spectrum for the anomalous Floquet topological insulator (AFTI) with parameters $(A,B,\Lambda) =(0.25,0.25,0)\omega$ and the staggered Floquet topological insulator (SFTI) with $(A,B,\Lambda) =(0.4,0.25,0.3)\omega$ in a ribbon geometry structure with zigzag edges (here, $b= |\bm b_\lambda|=\sqrt{3}a$). (b) The time evolution of the particle density distribution at the edge chain along the $x$-direction, $\rho_{\text{Edge}}(x, t)$, and the total density distribution, $\rho_{\text{Total}}(x, t) = \sum_y \rho (x, y, t)$, starting from an initial wave packet localized on a single ``red'' site at the edge. Parameters are the same as for the figures in (a).}
\label{figsm1}
\end{figure}

To visualize the difference between SFTI and AFTI, we consider a ribbon geometry with zigzag edges. In Fig.\ref{figsm1}(a), we respectively plot the spectrum for the AFTI with parameters $(A,B,\Lambda) =(0.25,0.25,0)\omega$ and the SFTI with parameters $(A,B,\Lambda) =(0.4,0.25,0.3)\omega$ (note that this parameter set with a relative large gap at $\epsilon=0$ can be smoothly connected to the SFTI phase region with $A=B$ shown in Fig.1(b) in the main text). As shown in Fig.\ref{figsm1}(b), we start with an initial wave packet localized at a red site at the edge and present the particle density distribution after time evolution for the two parameter choices. From the density distribution at the edge along the $x$-direction, $\rho_{\text{Edge}}(x)$, we observe obvious chirality in the wave packet's time evolution for AFTI and no chirality for SFTI. Note that even though the two edge states in SFTI have opposite chirality, they are indeed stable since the large energy difference ($\sim \omega/2$) between the two states forbids backscattering by weak disorder or interatomic interaction.

\section{II. Low-energy effective theory}

\noindent

In the following, we focus on the case $\Lambda=0$ and develop the low-energy effective model of the system. The approach can be generalized to the cases with $\Lambda\neq 0$. First, we apply a $\bf k$-independent unitary transformation, $U= \otimes_{m} \exp[-i\frac{\pi}{4}\sigma_{y}]$, to the
Hamiltonian in the extended Floquet Hilbert space, \mbox{$\mathcal{H} \rightarrow U \mathcal{H} U^\dagger$}. It means that each block matrix is rotated, $\mathcal{H}_{mn} \rightarrow\exp[-i\frac{\pi}{4}\sigma_y] \mathcal{H}_{mn} \exp[i\frac{\pi}{4}\sigma_y]$. Using this rotation, the Hamiltonian can be brought to a  simple form. Note that \begin{eqnarray}
\exp[-i\frac{\pi}{4}\sigma_y] \sigma_x \exp[i\frac{\pi}{4}\sigma_y] &=& -\sigma_z, \\
\exp[-i\frac{\pi}{4}\sigma_y] \sigma_z \exp[i\frac{\pi}{4}\sigma_y] &=& \sigma_x,
\end{eqnarray}
and $\sigma_y$ is invariant under the rotation. As a result, around the $\Gamma$ point, we have the transformed Hamiltonian
\begin{eqnarray}
 \mathcal{H}_{mm}({\bf k})
 &=& m\omega \sigma_0  +  A  \sum_{\lambda=1}^3 \left[\sigma_z \cos({\bf k}\cdot {\bm a}_\lambda)   -  \sigma_y  \sin({\bf k}\cdot {\bm a}_\lambda) \right] \notag\\
 &\simeq& m\omega \sigma_0  + z_0 \sigma_z ,
\end{eqnarray}
where $z_0 = 3 A -\frac{3}{4} A a^2 {\bf k}^2$ and $a= |{\bm a}_\lambda|$. Similarly,
\begin{eqnarray}
  \mathcal{H}_{m,m - 1}({\bf k})
 &=& \frac{ B}{2}  \sum_{\lambda=1}^3 e^{  i 2\pi \lambda/3} \left[\sigma_z  \cos({\bf k}\cdot {\bm a}_\lambda) - \sigma_y \sin({\bf k}\cdot {\bm a}_\lambda) \right]\notag\\
  &\simeq& z_1\sigma_z -i\sigma_y z_2.
\end{eqnarray}
where $z_1 =-{3 B a^2}(1- i\sqrt{3}) (k_x-ik_y)^2/{32}$ and $z_2 = {3 Ba}  (1- i\sqrt{3})(k_x+ik_y)/{8}$.

For the phase transition between FTI $(-1,0)$ and AFTI, the band crossing occurs at $\epsilon=\omega/2$. When $\omega$ is slightly different from $6A$ (the phase transition occurs at $\omega = 6 A$), we consider the low-energy effective Hamiltonian around the energy ${\omega}/{2}$, which is given by
\begin{equation}  \label{effH1}
\begin{split}
   & \mathcal{H}({\bf k}) -\frac{\omega}{2} = \\ & ~~ \begin{array}{c}
\begin{array}{ccccc}
& ~~~~m=1~~~~~~ ~~~ &~~~~~~~m=0~~~ &~
\end{array}\\
\left(\begin{array}{c|cc|cc|c}
\ddots  & \cdots & \cdots & \cdots &  \cdots & \cdots\\
\hline
\vdots &  {\omega}/{2} + z_0 & 0 & z_1 & -z_2
 & \cdots\\
\vdots &  0  & \color{red}{ {\omega}/{2}-z_0}
 & \color{red}{z_2} & -z_1  & \cdots\\
 \hline
\vdots & z^*_1 & \color{red}{z^*_2} &  \color{red}{-{\omega}/{2} +z_0} & 0 & \cdots\\
\vdots & -z^*_2 & -z^*_1 & 0& -{\omega}/{2} -z_0 & \cdots\\
\hline
\vdots  & \cdots & \cdots &  \cdots &  \cdots &  \ddots\\
\end{array}
\right)
\end{array}.
\end{split}
\end{equation}
Since $z_1$ and $z_2$ are small when ${\bf k} $ is small, we can omit them in the first step, and then the Hamiltonian is diagonalized. Because $\omega$ is close to $6A$, $\pm({\omega}/{2}-z_0)$ is close to zero, and other diagonal elements are of order $\omega$. For the low-energy approximation, we only keep the red elements as shown in Eq.\eqref{effH1}, and finally obtain
\begin{equation} \label{hamrd1}
    \mathcal{H}({\bf k})-\frac{\omega}{2} \simeq \begin{pmatrix}
    M_0& M_1 \\
    M_1^* & -M_0
    \end{pmatrix}
\end{equation}
where
\begin{eqnarray}
    M_0({\bf k}) &=& \frac{\omega}{2} -3A + \frac{3}{4}A a^2 {\bf k}^2,\\
M_1({\bf k}) &=& \frac{3 B a}{8} (1 -\sqrt{3}i) (k_x+ik_y),
\end{eqnarray}
when the effective mass $\mathcal{M}=\frac{\omega}{2} -3A$ is slightly larger than 0 (i.e., $\omega>6 A$). The topological phase transition happens when the effective mass vanishes (i.e., the gap $2\mathcal{M}$ vanishes). The Hamiltonian is thus simplified to a two-level model, and all high-energy bands are omitted. The self-energy of the Hamiltonian, according to Eq.(9) in the main text, has the following form
\begin{equation}
    \Sigma \simeq \delta \tau_z,\\
\end{equation}
where $\tau_z$ is the Pauli matrix in the low-energy two-level space,
\begin{equation}
    \delta =
    -\frac{W^2\sigma^2}{2\pi}\int d{\bf k} e^{- \frac{\sigma^2|{\bf k}|^2}{2}} \frac{M_0}{{M}_t^2}<0
\end{equation}
and
\begin{equation}
{M}_t^2 = M_0^2 + \frac{9 B^2}{16} a^2 {\bf k}^2.
\end{equation}
The off-diagonal self-energy correction vanishes since $M_1$ is an odd function of $\bf k$. The disorder suppresses the effective mass $\overline{\mathcal{M}} =\mathcal{M} +\delta$, where $\mathcal{M}>0$ and $\delta<0$, and thus the system goes into the AFTI phase when the disorder strength is increased.\\

\section{III. Effective parameters in the Born approximation}

\begin{table}[b]
\vspace{0.5cm}
  \centering
  \begin{tabular}{|c|c|c|c|c|}
  \hline
  \hline
       & $\delta\Lambda$ & $\delta \omega$ & $\delta A$    & $\delta B$ \\
\hline
$P_1$  & $0$             & $-5.42 W^2$     & $1.01 W^2 F $ & $-1.37 W^2 F$  \\
\hline
$P'_1$ & $0.18 W^2$      & $-5.44 W^2$     & $1.01 W^2 F $ & $\sim -1.39 W^2 F$ \\
\hline
$P_2$  & $0$             & $-1.70 W^2$     & $-0.91 W^2 F$ & $-0.12 W^2 F$ \\
\hline
$P_3$  & $-1.04 W^2$     & $-1.54 W^2$     & $-1.05 W^2 F$ & $\sim 0.08 W^2 F$ \\
\hline
\hline
\end{tabular}
\caption{The corrections to the parameters $\Lambda$, $\omega$, $A$, and $B$ due to the presence of disorder in the Born approximation. Here, $\delta \omega = (1/N)\sum_{i \in \text{all~sites}}(\Sigma_{11}^{ii}-\Sigma_{00}^{ii})$, $\delta \Lambda = (1/N)\sum_{i \in \text{blue~sites}, j \in \text{red~sites}}(\Sigma_{00}^{jj}-\Sigma_{00}^{ii})$, and $\delta A = -(1/3 N)\sum_{\langle i, j\rangle}\Sigma_{00}^{ij}$, where $N$ represents the total number of sites. The correction to the parameter $B$ can be extracted from the self-energy $\Sigma_{10}^{ij}$ for neighboring $i$ and $j$. For $P_1$ and $P'_1$, the reference energy in Eq.\,\eqref{selfx} is $\epsilon = \omega/2$ and for the others, $\epsilon = 0$. The correction to the parameter $B$ splits for a finite $\Lambda$, i.e., $\Sigma_{10}^{ij} \neq \Sigma_{10}^{ji}$ for two neighboring sites. The resulting $\delta B$ shown in the table for $P'_1$ and $P_3$ represents an average value, i.e., $\delta B =  -(2/3 N)\sum_{\langle i,j \rangle} \exp(-i \phi_\lambda) \Sigma_{10}^{ij}$ (each term in the summation gives the same value when $\Lambda =0$). Here, $F =\exp(-a^2/2\sigma^2)$.}\label{table1}
\end{table}

\begin{figure}[h]
  \centering
  \includegraphics[width=0.78\columnwidth]{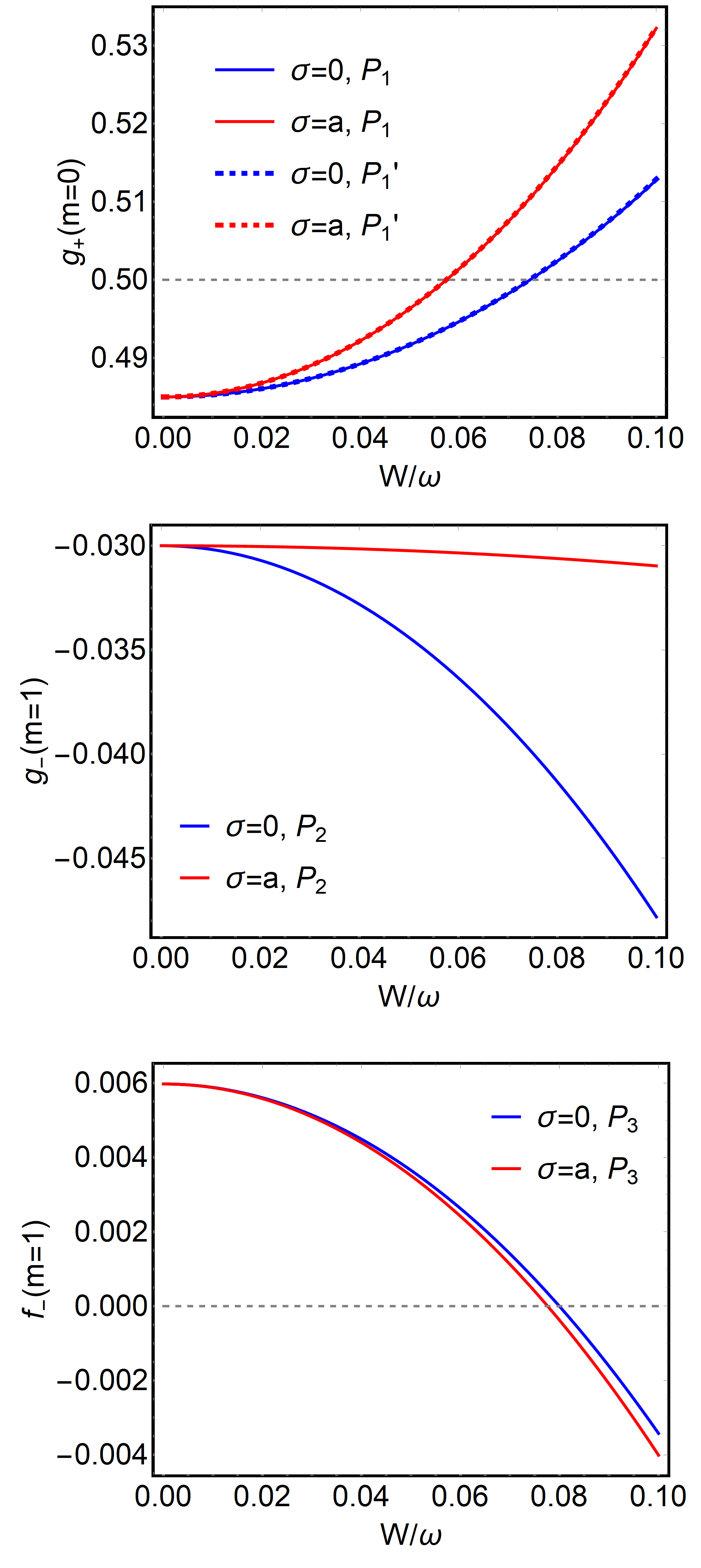}\\
  \caption{Topological phase transition induced by disorder in the Born approximation. The results for $P_1$, $P'_1$ and $P_3$ agree with the results obtained by calculating the disorder-averaged Bott index. However, there is no phase transition at $P_2$ in the Born approximation when the disorder strength is increased.}\label{figsm2}
\end{figure}

For Eq.\,(7) in the main text,
\begin{equation}\label{selfx}
    \Sigma_{mn}^{ij}(\epsilon) = G_{mn}^{ij}(\epsilon) \langle V_i V_j  \rangle,
\end{equation}
we calculate the corrections to the parameters $\Lambda$, $\omega$, $A$, and $B$ from disorder. For simplicity, we use the approximation $G(\epsilon) \simeq 1/(\epsilon-\mathcal{H})$ in the above equation. The corrections to these parameters for the Hamiltonian at different points $P_1$, $P'_1$, $P_2$ and $P_3$ are shown in Table\,\ref{table1}. From Eqs.(4) and (5) in the main text, we define the following two functions
\begin{equation}\label{gf}
{g}_{\pm}(m) = m \pm \sqrt{\left(\frac{\Lambda}{\omega}\right)^2 +9 \left(\frac{A}{\omega}\right)^2},
\end{equation}
and
\begin{equation} \label{gf2}
f_{\pm}(m) =\left(m-\frac{1}{2}\right) \pm \sqrt{\left(\frac{1}{2}\pm \frac{\Lambda}{\omega}\right)^2 + \frac{9 B^2}{4\omega^2}}.
\end{equation}
The topological phase transition occurs when either of these two functions crosses $0$ or $1/2$. Figure \ref{figsm2} shows how these values change after the correction of parameters. For $P_2$, the Born approximation shows no phase transition when increasing the disorder strength, which is inconsistent with the result from the disorder-averaged Bott index.

\end{document}